\documentclass[prd,reprint,preprintnumbers,longbibliography,nofootinbib]{revtex4-2}

\usepackage{multirow}
\usepackage{siunitx}

\usepackage{tikz}
\usetikzlibrary{decorations.markings, arrows, positioning, calc}

\usepackage{graphicx} 
\usepackage{latexsym}
\usepackage{mathtools}

\usepackage{amsmath,amssymb,amsbsy,amsfonts}
\usepackage{array}
\usepackage{bm}

\usepackage{graphics}
\graphicspath{{./fig/}{./tikz/}}

\usepackage{mathrsfs}
\usepackage{xcolor}
\usepackage{cancel}
\usepackage[normalem]{ulem}
\usepackage{enumerate}

\usepackage[unicode, pdfprintscaling=None, colorlinks]{hyperref}
\hypersetup{
    colorlinks,
    allcolors=blue!50!black,
}
\usepackage[capitalise]{cleveref}

\usepackage[scaled=0.8]{FiraMono}

\newcommand{\CC}{{\mathbb{C}}}

\renewcommand\>{\rangle}
\newcommand\<{\langle}

\newcommand\Order{O}

\DeclareMathOperator\Tr{Tr}

\usepackage{relsize}

\newcommand\xdag{{\vphantom{\dagger}}}

\usepackage{microtype}

\usepackage{xfrac}
\usepackage{orcidlink}
\newcommand\SU{\ensuremath{\text{SU}}}
\newcommand\CP{\ensuremath{\text{CP}}}
\newcommand\SO{\ensuremath{\text{SO}}}

\usepackage{ulem} 
\usepackage{amssymb} 

\newcommand\Lx{\ensuremath{L_X}}
\newcommand\Ly{\ensuremath{L_Y}}

\newcommand\del\partial

\newcommand\Gr{\ensuremath{\text{Gr}}}
\def\U{\ensuremath{\text{U}}}
\newcommand\PSU{\ensuremath{\text{PSU}}}

\newcommand\Sbar{\bar{S}}

\DeclareMathOperator*{\tint}{{\textstyle \int}}
\DeclareMathOperator*{\tsum}{{\textstyle \sum}}
\DeclareMathOperator*{\tprod}{{\textstyle \prod}}

\renewcommand{\lbrack}{\bm{[}}
\renewcommand{\rbrack}{\bm{]}}

\usepackage{microtype}
\usepackage{scalerel}

\newcommand\plaq{p}

\usepackage{cancel}
\usepackage{ulem}

\newcommand{\grfield}{\phi}

\newcommand\Nc{p}
\newcommand\Nr{k}

\usepackage{textgreek}
\usepackage{empheq}

\renewcommand\O{\ensuremath{\text{O}}}

\newcommand\ZZ{\mathbb{Z}}

\usepackage{diffcoeff}

\newcommand\gr{\Gr_\Nr(N)}
\newcommand\Id{\mathbb{I}}

\newcommand\SLa{\Lambda}
\newcommand\SLb{\bar{\Lambda}}

\newcommand\Csym{\ensuremath{\text{C}}}

\usepackage{relsize}

\usepackage[acronyms, smallcaps, nohypertypes={acronym}, nopostdot, style=super, nonumberlist, toc]{glossaries}
\glsenableentrycount
\makeglossaries  
\setacronymstyle{long-sc-short}

\newcommand\ac[1]{\gls{#1}}

\newcommand\acp[1]{\glspl{#1}}

\newacronym{WF}{wf}{Wilson-Fisher}
\newacronym{AF}{af}{asymptotically free}
\newacronym{RG}{rg}{renormalization group}
\newacronym{WZW}{wzw}{Wess-Zumino-Witten}
\newacronym[longplural={conformal field theories}]{CFT}{cft}{conformal field theory}
\newacronym[longplural={lattice field theories}]{LFT}{lft}{lattice field theory}
\newacronym[longplural={effective field theories}]{EFT}{eft}{effective field theory}
\newacronym[longplural={quantum field theories}]{QFT}{qft}{quantum field theory}
\newacronym{LEC}{lec}{low-energy constant}
\newacronym{QCD}{qcd}{quantum chromodynamics}
\newacronym{MC}{mc}{Monte Carlo}
\newacronym{IR}{ir}{infrared}
\newacronym{UV}{uv}{ultraviolet}
\newacronym{SNR}{snr}{signal-to-noise ratio}
\newacronym{NLSM}{nl$\sigma$m}{nonlinear sigma model}
\newacronym{PCM}{pcm}{principal chiral model}
\newacronym{CSA}{csa}{Cartan subalgebra}
\newacronym{SSB}{ssb}{spontaneous symmetry breaking}
\newacronym{DOF}{dof}{degrees of freedom}
\newacronym{DMRG}{dmrg}{densiy matrix renormalization group}
\newacronym{YM}{ym}{Yang-Mills}
\newacronym{QLM}{qlm}{quantum link model}
\newacronym{KG}{kg}{Kogut-Susskind}
\newacronym{KG-QLM}{kg-qlm}{Kogut-Susskind quantum link model}
\newacronym{D-QLM}{d-qlm}{D-theory quantum link model}
\newacronym{SPT}{spt}{symmetry-protected topological}
\newacronym{LSMOH}{lsmoh}{Lieb--Schultz--Mattis--Oshikawa--Hastings}

\begin{document}

\title{
  Lattice regularizations of $\theta$ vacua: Anomalies and qubit models
}

\author{Mendel Nguyen\,\orcidlink{0000-0002-7976-426X}}
\email{mendelnguyen@gmail.com}
\affiliation{Department of Physics, North Carolina State University, Raleigh, North Carolina 27607, USA}

\author{Hersh Singh\,\orcidlink{0000-0002-2002-6959}}
\email{hershsg@uw.edu}
\affiliation{InQubator for Quantum Simulation (IQuS), Department of Physics, University of Washington, Seattle, Washington 98195-1550, USA}
\affiliation{Institute for Nuclear Theory, University of Washington, Seattle, Washington 98195-1550, USA}

\preprint{IQuS@UW-21-031, INT-PUB-22-027}

\begin{abstract} Anomalies are a powerful way to gain insight into possible lattice regularizations of a quantum field theory. In this work, we argue that the continuum anomaly for a given symmetry can be matched by a manifestly-symmetric, local, lattice regularization in the same spacetime dimensionality only if (i) the symmetry action is offsite, or (ii) if the continuum anomaly is reproduced exactly on the lattice. We consider lattice regularizations of a class of prototype models of QCD: the (1+1)-dimensional asymptotically-free Grassmannian \acp{NLSM} with a $\theta$ term. Using the Grassmannian \acp{NLSM} as a case study, we provide examples of lattice regularizations in which both possibilities are realized. For possibility (i), we argue that Grassmannian \acp{NLSM} can be obtained from $\SU(N)$ antiferromagnets with a well-defined continuum limit, reproducing both the infrared physics of $\theta$ vacua and the ultraviolet physics of asymptotic freedom. These results enable the application of new classical algorithms to lattice Monte Carlo studies of these quantum field theories, and provide a viable realization suited for their quantum simulation. On the other hand, we show that, perhaps surprisingly, the conventional lattice regularization of $\theta$ vacua due to Berg and L\"uscher reproduces the anomaly exactly on the lattice, providing a realization of the second possibility.

 \end{abstract}

\maketitle
\glsresetall

\section{Introduction}
\label{sec:introduction}

  The Standard Model of particle physics, and in particular the strong sector described by \ac{QCD}, has many physically relevant yet so-far inaccessible regimes beyond perturbation theory.
While non-perturbative methods such as lattice field theory using classical \ac{MC} methods have been used with remarkable success, many problems of interest such as real-time dynamics, finite-density or nontrivial $\theta$ vacua remain inaccessible.
Recently, the growth of quantum technologies has opened up the possibility of exploring these questions.
However, the infinite-dimensional local Hilbert space of standard bosonic lattice field theories limits their applicability to near-term quantum platforms with low qubit counts.
This has motivated a search for unconventional lattice regularizations of \acp{QFT} with finite-dimensional local Hilbert spaces.
New lattice regularizations of a given \ac{QFT} can also provide solutions to sign problems and enable application of cluster algorithms.
Success in low-dimensional asymptotically-free prototypes of \ac{QCD} establishes a route towards developing new tools for lattice studies of \ac{QCD}, both on classical and quantum hardware
\cite{jordan_quantum_2014, jordan2012quantum, yeter2019scalar, klco2019digitization, chandrasekharan_spin_2002, brower_dtheory_2004, beard_efficient_2006, laflamme_cp_2016, evans_su_2018a, bruckmann_nonlinear_2019, nuqscollaboration_ensuremath_2019, alexandru_universality_2021, alexandru_spectrum_2022, chandrasekharan_quantum_1997a, brower_qcd_1999, raychowdhury2020solving, anishetty_prepotential_2009, banerjee2013atomic, zohar2015quantum, banuls2017efficient, muschik2017u, zache2018quantum, alexandru2019gluon, bender2020gauge, davoudi2020towards, klco20202, shaw2020quantum, kasper2020non, buser2021quantum, haase2021resource, alexandru_spectrum_2022, caspar_asymptotic_2022, singh_qubit_2022a, zhou_spacetime_2022, bhattacharya_qubit_2021, singh_qubit_2019}.

If one goes beyond na\"ive discretizations of the continuum Hamiltonian (and perturbative improvements thereof), then the space of possible lattice regularizations of a given \ac{QFT} is immense.
In general, a lattice regularization for a given \ac{QFT} is any lattice Hamiltonian (or action) which has the correct quantum critical point and relevant parameter (in the \ac{RG} sense).
We might ask: how do we find new lattice regularizations, perhaps with unique advantages for quantum and classical simulations? 
While it is clear that symmetries are an important part of this story,
the point of view of this work is that this search can be refined by considerations of anomalies.

To make this concrete, in this work, we consider lattice regularizations of a certain class of (1+1)-dimensional \acp{NLSM} with a $\theta$ term, called the Grassmannian $\gr$ \acp{NLSM},
which are well-known prototypes of \ac{QCD}.
These are \acp{NLSM} with the target space
\begin{align}
\gr = \U(N) / \lbrack \U(k) \times \U(N-k)\rbrack,
\end{align}
formally defined by the continuum action
\begin{align}
  S = \frac{1}{g^2} \int d^2 x  \Tr (\del_{\mu} P )^2 
  + \frac{\theta}{4\pi} \int d^2x\, \epsilon^{\mu \nu} \Tr P\, \del_\mu P\, \del_\nu P
    \label{eq:S-nlsm}
\end{align}
where $P(t,x) \in \Gr_k(N)$ is an $N\times N$ Hermitian projector matrix such that $P^2 = P = P^\dagger$ and $\Tr P = \Nr$.
Important examples of this class are the $k=1$ models, which are also known as $\CP(N-1)$ models.
All these models are asymptotically-free, have a dynamically generated mass scale, and admit a $\theta$ term. The $\Gr_1(2)=\CP(1)$ case, also called the $\O(3)$ \ac{NLSM}, has been particularly important as a testbed for developments in lattice \ac{QCD}.

A majority of the lattice work in understanding these models has used the lattice formulation which is a direct discretization of \cref{eq:S-nlsm}, which we refer to as the ``conventional'' lattice regularization. 
In this regularization, a satisfactory topological definition of the $\theta$ term was proposed by Berg--L\"uscher for the $\CP(N-1)$ model.
While it has nice topological properties, a lattice \ac{MC} study of general $\theta$ vacua has been limited by a severe sign problem at $\theta\neq0$, except in special cases.
(It can be somewhat alleviated for the $\O(3)$ model using a meron cluster formulation \cite{bietenholz_testing_1996, deforcrand_walking_2012, bogli_nontrivial_2012}.)

Recently, another type of lattice regularization was proposed: as an antiferromagnetic model of qubits \cite{caspar_asymptotic_2022}, such that a controlled continuum limit  at arbitrary $\theta$ can be taken by adding a small extra dimension. We call this a ``qubit'' regularization.
Not only is this a very natural regularization of the $\theta$ vacua on quantum computers, it also does not suffer from a sign problem at nontrivial $\theta$ for classical \ac{MC} computations.
However, it does not have manifest topological properties, which only become apparent as one takes the continuum limit.
Therefore, both these types of regularizations have their own advantages.

In this work, we take a step back and attempt to understand these two types of regularizations from the point of view of anomalies.  In particular, these models have an 't~Hooft anomaly at $\theta=\pi$ \cite{Gaiotto:2017yup}, which presents obstructions for certain kinds of symmetric lattice regularizations.  The obstruction from the anomaly can manifest on the lattice in two ways, and we argue that the two above-mentioned regulators correspond to precisely these two ways.  

The presence of an anomaly for a given symmetry $G$ implies that for a $G$-symmetric lattice regularization, the symmetry cannot be both \emph{manifest} and \emph{onsite}.
(We give a more detailed argument and explain our definitions in \cref{sec:overview}.)
A $G$-anomaly in the continuum theory implies that it should be impossible to gauge $G$ -- this should also be true on the lattice.
If the symmetry is offsite, then it cannot be gauged in the usual manner, and therefore the constraint is satisfied.  On the other hand, if the symmetry is onsite, 
then the only way for the lattice regularization to match the anomaly is to explicitly reproduce the anomaly.

We can now place the two regularizations of the $\CP(1)$ or the $\O(3)$ model in this framework.
Here the anomaly is for $G = \SO(3) \times \Csym$, where $\Csym$ is charge conjugation.
This particular anomaly is especially interesting since an analogous anomaly exists for 3+1-dimensional $\SU(N)$ Yang-Mills at $\theta=\pi$ \cite{Gaiotto:2017yup}.

The qubit regularization of the $\theta$ term is of the first kind.
Starting from the theory at $\theta=\pi$, $\Csym$ is realized offsite as a translation-by-one symmetry.
Therefore, we can tune $\theta$ away from $\pi$ by simply breaking the translation symmetry, which can be done by introducing staggered couplings.
This generates a $\theta$ term for the low-energy effective theory.

On the other hand, the conventional regularization with the Berg--L\"uscher $\theta$ term in fact does have an exact $G$ symmetry with onsite action.
As argued above, this should imply that this regularization exhibits the anomaly exactly on the lattice. As we show in this work (see \cref{sec:luscher}), this is indeed the case. This is perhaps surprising since it is often assumed that there are no anomalies on the lattice. 
A gauging procedure, much like the one used to derive the anomalies in the continuum, works for the lattice theory and reproduces the anomaly.
In fact, our derivation can also be thought of as an independent computation of the anomaly, using a well-defined lattice model.
Therefore, this well-known regularization is an example of the second possibility.

More recently, another illuminating perspective has come in the language of \ac{SPT} phases.
't Hooft anomalies in a $D$-dimensional (Euclidean) theory correspond to \ac{SPT} phases of a $(D+1)$-dimensional theory, which is the so-called bulk/boundary correspondence \cite{wen_classifying_2013}.  From this point of view, a $D$-dimensional theory with an 't Hooft anomaly can naturally arise on the boundary of a $(D+1)$-dimensional nontrivial \ac{SPT} phase. However, there are obstructions when trying to formulate theory in a manifestly $D$-dimensional setup.
Refs.~\cite{kravec_gauge_2013,kravec_allfermion_2015,wang_nonperturbative_2022} have attempted to formulate the Nielsen--Ninomiya theorem in the language of \ac{SPT} phases.
Their claim is that it is impossible to obtain a local, symmetric, manifestly $D$-dimensional lattice regularization of the boundary theory.
However, as we argue, their conclusions are somewhat incomplete. It is in fact \emph{not} impossible to obtain a manifestly $D$ dimensional local realization of the boundary of an \ac{SPT} phase.
The ``no-go'' theorem can be avoided in a rather direct way: the lattice theory can manifestly exhibit the anomaly.

In the condensed matter literature, some of this has been long appreciated in the context of \ac{LSMOH} theorems.  In particular, the connection between \ac{LSMOH} and anomalies has been delineated in Refs.~\cite{jian_liebschultzmattis_2018, cho_anomaly_2017}.
It is pointed out in Ref.~\cite{cho_anomaly_2017}, that if a $D$-dimensional (Euclidean) continuum theory with a $G$-anomaly is known to arise as the low-energy effective theory of a $D$ dimensional lattice model, then either (i) $G$ must be offsite, or (ii) $G$ must not be an exact symmetry of the lattice model.
This is sometimes phrased as a ``no-go'' theorem.
Here we show that there is another possibility which is often missed: the symmetry is both onsite and exact, but the lattice theory explicitly has the anomaly.

There have been several other recent examples of anomalies on the lattice. For example, this has been shown for K\"ahler--Dirac fermions on triangulated lattices
\cite{catterall_topological_2018, catterall_induced_2022, catterall_ahlerdirac_2018, butt_anomalies_2021, catterall_hooft_2022}.
In the case of chiral fermions, formulations satisfying the Ginsparg--Wilson relation have been shown to reproduce the anomaly on the lattice for a modified chiral symmetry 
\cite{neuberger_exactly_1998, luscher_exact_1998, hasenfratz_index_1998,kaplan_chiral_2012}.

This manuscript is organized as follows. In \cref{sec:overview}, we present our general arguments regarding lattice regularizations for theories with anomalous symmetries, and present an overview of the two examples of lattice regularizations which we consider in this work, explaining how they realize the anomalies in different ways.
Then in \cref{sec:dtheory}, we discuss a ``qubit'' regularization, with finite-dimensional local Hilbert spaces, of the Grassmannian nonlinear sigma models.
Then in \cref{sec:luscher}, we discuss lattice regularizations exhibiting the anomalies explicitly, including a generalization of the well-known regularization of the $\theta$-term in $\CP(N-1)$ models due to Berg--L\"uscher to the $\gr$ models. 
Finally, in \cref{sec:conclusion}, we summarize our results and comment on some future directions.


\section{Obstructions to lattice regularizations from anomalies}
\label{sec:overview}

One of the aims of this work is to emphasize how obstructions to regulating a theory on the lattice arising from anomalies guide us in constructing new lattice regularizations. 

In this work, we always refer to ``anomaly'' in the sense of an \emph{'t~Hooft anomaly}: a symmetry $G$ is said to be 't~Hooft anomalous if it cannot be consistently gauged.
We emphasize that $G$ is a genuine symmetry of the theory, and that the presence of an 't~Hooft anomaly does not indicate an inconsistency of the theory.
We shall also confine our attention to \emph{internal} symmetries.

In many cases of interest, the $G$ anomaly has a bit more structure.
If $G$ is generated by two subgroups $G_1,G_2$ such that there is no obstruction to gauging $G_1$ or $G_2$ individually,
but gauging $G_1$ breaks $G_2$, or vice versa, then we say there is a \emph{mixed} anomaly between $G_1$ and $G_2$.
In other words, $G_1$ and $G_2$ cannot be gauged simultaneously.
In all the cases we consider in this work, we always have a mixed anomaly, although the arguments apply more generally.

From the point of view of lattice regularizations, the presence of an 't~Hooft anomaly shows up as an obstruction to constructing certain type of lattice regularizations:
an anomaly in $G$ for a $D$-dimensional continuum theory implies that we cannot have a $D$-dimensional $G$-symmetric lattice regularization with $G$ both \emph{manifest} and \emph{onsite}.

  By a ``manifest'' symmetry, we mean an exact symmetry of the theory which is also an invariance of the action.
  In contrast, a nonmanifest symmetry is one where the action is noninvariant, but all correlation functions are invariant.
  We emphasize that in both cases, we are referring to an exact symmetry of the theory.

 By an ``onsite'' symmetry, we mean that the symmetry group acts on the full Hilbert space as a tensor product of representations on the local Hilbert spaces.
(In bosonic field theories, the full Hilbert space of a lattice model naturally admits a factorization as a tensor product of Hilbert spaces at each lattice site, which we call the \emph{local} Hilbert space.)
  We shall also say that a non-onsite symmetry is ``offsite.''

  If the symmetry $G$ were manifest and onsite, then we could trivially gauge them both on the lattice by introducing appropriate link variables, and this would violate the anomaly.
Therefore, a local $G$-symmetric $D$-dimensional lattice regulator for a theory with a $G$ anomaly must
\begin{enumerate}[(i)]
\item realize $G$ offsite, or
\item realize $G$ onsite but nonmanifestly.
\end{enumerate}

Another widely-appreciated possibility is that the $D$-dimensional theory arises as the boundary theory of a $(D+1)$-dimensional lattice model in an \ac{SPT} phase \cite{wen_classifying_2013, wang_nonperturbative_2022, kaplan_method_1992}.
In such cases, it is possible to obtain a lattice regulator with an exact onsite symmetry, albeit at the cost of introducing an extra dimension.
Yet another possibility is to consider lattice regulators in which $G$ is not a symmetry microscopically but arises at large distances.
In this work, we only consider symmetric lattice regulators in $D$ spacetime dimensions, in which case we have only the two possibilities listed above.

As mentioned in the introduction, while possibility (i) is well-known, especially in the context of \ac{LSMOH} theorem in condensed-matter literature \cite{jian_liebschultzmattis_2018, cho_anomaly_2017},  possibility (ii) is often missed.
Indeed, the anomaly implies that there must somehow be an obstruction to gauging $G$, while if $G$ acts onsite, there seems to be no obstruction to turning on a background gauge field for it. The only possibility, therefore, seems to be that the dependence on the background gauge field cannot be gauge invariant, which is to say that the lattice regulator must
\begin{itemize}
    \item[(ii*)] \emph{explicitly exhibit the anomaly}.
\end{itemize}
Indeed, this will be the case in the examples discussed in \cref{sec:luscher}. It is also the case in the modified Villain models \cite{Sulejmanpasic:2019ytl, gorantla_modified_2021, Anosova:2022cjm, Anosova:2022yqx}.
We should also note that it may be possible for a lattice regulator with offsite symmetry to also explicitly exhibit an anomaly, such as Ginsparg--Wilson fermions \cite{luscher_exact_1998,wang_nonperturbative_2022}.

In this work, we provide examples for both scenarios (i) and (ii), using the $(1+1)$-dimensional $\gr$ \ac{NLSM} as a case study.

\subsection{$\O(3)$ nonlinear sigma model}

To simplify the discussion a bit, let us first consider the case of $\Gr_1(2) = \CP(1)$, or the $\O(3)$ model with a $\theta$ term. This model can be formulated in terms of a unit 3-vector field $\vec n(x)$ with Euclidean action
\begin{align}
  S = \frac{1}{2g^2}\int d^2x\, (\del_\mu \vec n)^2 + \frac{i\theta}{8\pi} \int d^2 x\, \epsilon^{\mu \nu} \vec n \cdot (\del_\mu \vec n \times \del_\nu \vec n).
\end{align}
At any $\theta$, the theory has a global $\SO(3)$ symmetry. At $\theta=0,\pi$ it also has a charge conjugation symmetry,
\begin{align}
   \Csym \colon\vec n(x) \mapsto -\vec n(x).
  \label{eq:C-O(3)}
\end{align}

At $\theta=\pi$, this theory has a mixed anomaly between \SO(3) and \Csym{} symmetries
\cite{Gaiotto:2017yup}.
The above argument then implies that a lattice regularization of the $\theta=\pi$ $\O(3)$ \ac{NLSM} must either realize $\SO(3) \times \Csym$ offsite, or it must explicitly exhibit the anomaly.

One lattice regularization of the $\O(3)$ \ac{NLSM} is provided by the spin-$\frac{1}{2}$ Heisenberg antiferromagnet
\begin{align}
  H = J \tsum_{\< i j \>} \vec S_{i} \cdot \vec S_{j},
\end{align}
where $\vec S_{i}$ are spin-$\frac{1}{2}$ operators and the sum runs over nearest neighbor sites $i,j$.
In one spatial dimension, this
is famously known to be described by the $\O(3)$  \ac{NLSM} at $\theta=\pi$ at low energies 
\cite{haldane_nonlinear_1983, haldane_continuum_1983, affleck_critical_1987}.
Interestingly, this can be thought of as not just a low-energy \ac{EFT} but also a genuine lattice regularization by adding a small extra dimension of size $L'$ (odd) in the D-theory formulation
\cite{chandrasekharan_spin_2002, chandrasekharan_quantum_1997a, brower_qcd_1999, brower_dtheory_2004}.
As shown in Refs.~\cite{beard_study_2005, caspar_asymptotic_2022}, the continuum limit of the $\O(3)$ model at $\theta=\pi$ can be obtained by taking $L' \to \infty$ (odd).
This not only reproduces the physics of $\theta=\pi$ in the \ac{IR}, but also the asymptotic freedom in the \ac{UV}.

In this regularization, the $\SO(3)$ symmetry is clearly manifest and onsite, and therefore can be gauged easily.  However, the continuum charge conjugation symmetry acts on the lattice as translation by one unit
\cite{affleck_field_1988}
\begin{align}
   \Csym \colon\vec S_{i} \mapsto \vec S_{i+1}
\end{align}
and therefore realizes possibility (i) of the above theorem. 

It is especially interesting to consider the case of a widely-used lattice regularization of the $O(3)$ \ac{NLSM} on a two-dimensional Euclidean spacetime lattice, given by the following action at $\theta=\pi$: 
\begin{align}
  S_0[\vec n] =  - \frac{1}{g^2} \tsum_{\< i j \>} \vec n_i \cdot \vec n_{j}
  \label{eq:S-O3}
\end{align}
where the $\vec n_i$ are real unit 3-vectors and the sum runs over nearest neighbor sites $i,j$ on a square lattice.
In the $g\to 0$ limit, this defines a lattice regularization for the $\theta=0$ theory.
This has a manifest onsite $G=\SO(3) \times \Csym$ symmetry, with $ \Csym \colon\vec n_i \mapsto - \vec n_i$.

Let us now assume that there is a local $G$ symmetric deformation of the action in \cref{eq:S-O3} which allows us to switch on $\theta=\pi$,
\begin{align}
  S_{\theta=\pi}\lbrack \vec n\rbrack = S_0\lbrack\vec n\rbrack + S_1\lbrack\vec n\rbrack.
\end{align}
But if this were the case, we would be able to gauge $G$, violating the constraint from the anomaly. Therefore, we conclude that it is impossible to obtain the $\theta=\pi$ from \cref{eq:S-O3} in this manner.

The above argument might at first glance lead one to think that there simply cannot be a lattice regularization of the $\theta=\pi$ theory with onsite $\SO(3) \times \Csym$ symmetry, but this would be too hasty. 
Indeed, there is a well-known construction of the $\theta$ term on the lattice for the $\CP(N-1)$ models, due to Berg and L\"uscher \cite{berg_definition_1981, luscher_topology_1982}, which is $\SO(3) \times \Csym$ symmetric at $\theta=0,\pi$.
In their formulation, the action of $\SO(3) \times \Csym$ symmetry is onsite.  
The reason their formulation avoids the ``no-go theorem'' is that while the Boltzmann weight $\exp(-S)$ is invariant under $\Csym$, the action itself is not. In this sense, the $\Csym$ invariance is not manifest in their formulation.
(A similar point has been made by Ref.~\cite{gorantla_modified_2021} in the context of modified Villain models.)  Indeed, the same is true in the continuum formulation \eqref{eq:S-O3}.
A ``no-go theorem'' for this case can be therefore be phrased as
\begin{quote}
There cannot be a lattice regulator for the $\theta=\pi$ $\O(3)$ \ac{NLSM} with \emph{manifest onsite} $\SO(3) \times \Csym$ symmetry.
\end{quote}

On the other hand, there is no obstruction to gauging $\SO(3)$.
And if $\SO(3) \times \Csym$ is an exact onsite symmetry of lattice theory such that $\SO(3)$ can be gauged, then the only way out is if the lattice theory explicitly produces the anomaly.
As we will show in \cref{sec:luscher}, this is indeed the case, giving us an example of possibility (ii) above. 

\subsection{Generalization to $\gr$ models}

The previous discussion about the $\O(3)$ model can be generalized neatly to the general class of asymptotically-free $\Gr_k(N)$ models with a $\theta$ term, given by the continuum action in \cref{eq:S-nlsm}.
The $P$ field transforms under the action of $\PSU(N)$ symmetry, as well as charge conjugation
\begin{align}
  \Csym \colon P(x) \mapsto P(x)^*.
\end{align}
$\PSU(N)$ symmetry is present at any $\theta$, while $\Csym$ symmetry is present only at $\theta=0,\pi$.
As discussed in \cref{sec:anomalies}, a continuum analysis of mixed anomalies between $\PSU(N)$ and $\Csym$ symmetries reveals \cite{Dunne:2018hog}:
\begin{enumerate}
\item $N=\text{even}$, $\Nr=\text{odd}$: At $\theta=\pi$, there is a mixed anomaly between $\PSU(N)$ and $\Csym$ symmetries.
\item Otherwise: There is a ``global inconsistency'' between the $\theta=0$ and $\theta=\pi$ theories.
\end{enumerate}
For the first case, we have an immediate generalization of the above argument for the $\O(3)$ model. However, the second case results in a more subtle type of obstruction to lattice regularization.

A ``global inconsistency''
\cite{kikuchi_global_2017,Gaiotto:2017yup,tanizaki_vacuum_2017}
between the $\theta=0,\pi$ continuum theories says that the action cannot be modified by a local symmetric counterterm such that both points are anomaly free. In other words, unlike the case $(N,k)$ = (even, odd), we can remove the anomaly at $\theta=\pi$, but at the cost of introducing an anomaly at $\theta=0$.  From the point of view of a lattice regularization, this suggests an interesting type of constraint.
While, in principle, it is possible to have lattice regulators with manifest onsite $\PSU(N) \rtimes \Csym$ symmetry at both $\theta=0,\pi$ points, we have the following obstruction: 
\begin{quote}
There cannot be a lattice regulator which allows one to \emph{continuously} tune $\theta$ from $0$ to $\pi$, such that at $\theta=0,\pi$ there is manifest onsite $\PSU(N) \rtimes \Csym$ symmetry.
\end{quote}

To see this, let us assume that we have a lattice action $S_0$ which furnishes a lattice regularization of the $\theta=0$ theory with a manifest onsite $\PSU(N) \rtimes \Csym$ symmetry.  Now, let us further assume that there is a local $\PSU(N)$ symmetric perturbation $K_\gamma$ parametrized by a continuously tunable parameter $\gamma\in [0,1]$, 
\begin{align}
  S_\gamma = S_0 + K_\gamma
\end{align}
The perturbation $K_\gamma$ is such that $K_0 = 0$, and $S_1$ furnishes a regularization for the $\theta=\pi$ theory.
If $K_1$ is invariant under the onsite $\Csym$ action, then we can gauge $\PSU(N)$ symmetry such that $\Csym$ symmetry survives at both $\theta=0,\pi$. This contradicts global inconsistency, which requires that $\Csym$ be explicitly broken at either $\theta=0$ or $\theta=\pi$.
Therefore, such a regulator cannot exist.

We present two kinds of regularizations in this work, generalizing the previous discussion of the $\CP(1)$ model to $\gr$ models. The first kind is based on $\SU(N)$ antiferromagnets, and realizes possibility (i), with an offsite implementation of the $\Csym$ symmetry. As in the $\CP(1)$ case
\cite{caspar_asymptotic_2022}, 
a $\theta$ term can be introduced by staggering the couplings, and continuum limit can be taken with the D-theory prescription.
We discuss the details in \cref{sec:dtheory}.

In the second kind, we consider models on Euclidean spacetime lattices, such as the conventional lattice regularization of the $\gr$ models, with the $\theta=0$ theory given by the lattice action 
\begin{align}
  S_{0}\lbrack P \rbrack = -\frac{1}{g^2} \tsum_{\< i j \>} \Tr P_{i} P_{j}
  \label{eq:S-gr}
\end{align}
where $P_i \in \gr $ are $N\times N$ Hermitian projector matrices with $P^2 = P$ and $\Tr{P} = k$.
We generalize the construction of Berg--L\"uscher and construct a lattice $\theta$ term for this model.
This model has an exact onsite $\PSU(N) \rtimes \Csym$ symmetry.
Therefore, this model must exhibit the anomaly explicitly on the lattice.  Using a method very analogous to the continuum, we indeed find that this model reproduces the continuum anomaly exactly, and therefore provides an example of possibility (ii) of the no-go theorem.
We provide details on the construction and the computation of the anomaly in \cref{sec:luscher}.


\section{A qubit regularization of Grassmannian models}
\label{sec:dtheory}

\begin{figure}[htb]
  \centering
  \includegraphics{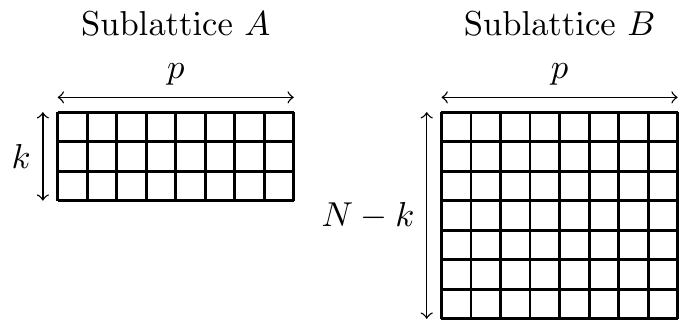}
  \caption{Young tableau for the representations each sublattice of the $\SU(N)$ Heisenberg antiferromagnet on a bipartite lattice. On the sublattice $A$, we choose the representation $R=(\Nr,\Nc)$ given by a Young tableau of $\Nc$ columns and $\Nr$ rows, while on the sublattice $B$, we choose the conjugate representation $\bar R = (N-\Nr, \Nc)$. }
\label{fig:sublattices}
\end{figure}

\begin{figure*}[htb]
  \includegraphics{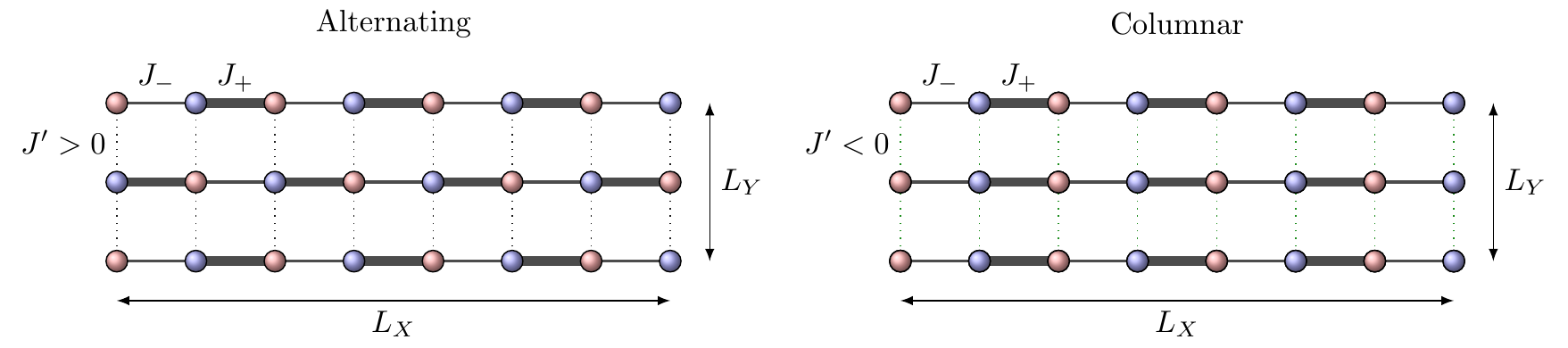}
  \caption{Viable lattice regularizations of the $\gr$ \ac{NLSM} with  a $\theta$ term using $\SU(N)$ antiferromagnets.  The red and blue dots indicate sites at different sublattices, with representations given by the rectangular Young tableaux shown in \cref{fig:sublattices}. The interactions within the same sublattice are always ferromagnetic, while interactions between different sublattices are always antiferromagnet.  The thickness of the bonds indicates alternating strength of the interactions.  The continuum limit of the $\gr$ \ac{NLSM} at nontrivial $\theta$ is obtained in the $\Ly \to \infty$ limit with fixed $\gamma \Ly$.}
  \label{fig:staggering}
\end{figure*}

In this section, we propose a lattice regularization for the Grassmannian nonlinear sigma models with a $\theta$ term using $\SU(N)$ antiferromagnets with staggered couplings, generalizing recent results for the $\CP(1)$ model \cite{caspar_asymptotic_2022}.

The model is defined on a two-dimensional $\Lx\times \Ly$ square bipartite lattice. We denote the two sublattices as $\SLa$ and $\SLb$. On sublattice $\SLa$, we place $\SU(N)$ spins in the representation $R=(\Nr, \Nc)$ given by the rectangular Young tableau with $\Nr$ rows and $\Nc$ columns. 
On sublattice $\SLb$, we place $\SU(N)$ spins in the conjugate representation $\bar R = (N-\Nr, \Nc)$, as shown in \cref{fig:sublattices}. Different choices of the bipartitions can be considered. We show two such choices in \cref{fig:staggering}.
Given a choice of the bipartition and representations, we write the nearest-neighbor $\SU(N)$ symmetric Heisenberg Hamiltonian with staggered couplings
\begin{align}
  H =
  \tsum_{(x,y)} J_{x,y}\ T_{x,y}^{\alpha}\ T_{x+1,y}^{\alpha}
  + J' \tsum_{ (x,y) } \ T_{x,y}^{\alpha}\ T_{x,y+1}^{\alpha}
  \label{eq:ham-SU(N)}
\end{align}
where $T_{\alpha}$ are the $\SU(N)$ generators ($\alpha=1,\dotsc,N^2-1$) in representation $R$ or $\bar{R}$ depending on the site $(x,y)$ ,
$J_{x,y}$ are the couplings along the $x$ direction, $J'$ is the coupling along the $y$ direction.
As shown in \cref{fig:staggering}, we consider the two configurations for staggering the couplings:
\begin{align}
  \text{Alternating:} && J' > 0, \quad J_{x,y} &= J\lbrack 1 + (-1)^{x+y} \gamma\rbrack,
  \label{eq:staggering} \\
  \text{Columnar:} && J' < 0, \quad J_{x,y} &= J\lbrack 1 + (-1)^{x} \gamma \rbrack,
\end{align}
where $J>0$ is always antiferromagnetic, and $\gamma$ is the staggering parameter.
Given a choice of bipartition $\SLa, \SLb$, the couplings $J, J', \gamma$ are chosen such that interactions within the sublattice are ferromagnetic, while the interactions between different sublattices are antiferromagnetic.

The representations $R, \bar{R}$ and the couplings are chosen such that the two-dimensional model is in the massless broken phase  in the $\Lx, \Ly \to  \infty$ limit.  (We discuss this in more detail below.)  With this choice of representations, a continuum limit can be defined on a $\Lx \times \Ly$ spatial lattice, in the regime $\Lx \gg \Ly \gg 1$.
More precisely, the continuum limit can be taken by considering $\Ly$ even or odd, and taking $\Ly \to \infty$ with fixed $\gamma \Ly$, resulting in the $\gr$ \ac{NLSM} with
\begin{align}
  \theta = \pi \Nc \Ly (1 + \gamma f)
\end{align}
where $f$ is a non-universal parameter.

\subsection*{Low-energy physics and $\theta$ vacua}

It was shown by Read and Sachdev \cite{read_features_1989} using coherent-state formalism that in the large-$\Nc$ limit, the two-dimensional $\SU(N)$ antiferromagnet with the above choice of representations results in a $\gr$ \ac{NLSM} at low energies.
For a one-dimensional $\SU(N)$ antiferromagnetic chain, it can also be easily shown that a $\theta$ term is induced at low energies by staggering the couplings (see \cref{sec:coherent-states}). This can be understood by the fact that the $\Csym$ symmetry is realized offsite as a translation by one site, 
and therefore a staggered coupling breaks the $\Csym$ symmetry, which should generically induce a $\theta$ term in the \ac{NLSM}.
For $\SU(N)$ ladders with $\Ly$ fixed, we can argue that the model is effectively a one-dimensional \ac{NLSM}
\cite{sierra_application_1997, sierra_nonlinear_1996} and similar considerations apply.  Indeed, in Refs.~\cite{sierra_application_1997, sierra_nonlinear_1996, martin-delgado_phase_1996} a semiclassical analysis in the large-$(S\Ly)$ limit was performed for the spin-$S$ $\SU(2)$ ladders with staggered couplings, and it was shown that the long-distance physics is that of a (1+1)-dimensional $O(3)$ \ac{NLSM} with $\theta = \pi S \Ly (1 + c \gamma)$, where $\gamma$ determines the strength of staggering and $c$ is a nonuniversal parameter.

As we have argued above, the low-energy physics of the $\SU(N)$ model of \cref{eq:ham-SU(N)} is given by the $\gr$ \ac{NLSM} with a $\theta$ term.
However, to claim that we have a lattice regularization of a \ac{QFT}, this is not enough -- we must have a prescription for taking the continuum limit. 

\subsection*{Continuum limit and asymptotic freedom}

An elegant prescription for taking the continuum-limit is provided by the \emph{D-theory} approach  
\cite{brower_qcd_1999, chandrasekharan_quantum_1997a, chandrasekharan_spin_2002}.  
In this approach, a $(d+1)$-dimensional asymptotically-free theory can be obtained from a system in $(d+1)+1$ spacetime dimensions, where the extra spatial dimension is taken to be small in physical units.
There are two ingredients for this to work.  The first one is that the $(d+1)$-dimensional model should be in a massless phase, and the second is dimensional reduction.

\newcommand\xiSR{\xi_{\mathrm{SR}}}

Let us assume that the two-dimensional model is in broken phase with massless Goldstone modes. (We will discuss the conditions for this in more detail below.)
We take the extent in the extra spatial dimension to be $\Ly$.
The low-energy continuum action (without the $\theta$ term) is
\newcommand\geff{g_{\text{eff}}}
\begin{align}
  S_{0} = \frac1{2g^2} \smashoperator{\int_0^{\Ly}} dy \int d x\, dt\,  \Tr \lbrack (\del_{x} P )^2 + (\del_{t} P )^2 + (\del_yP)^2 \rbrack 
\end{align}
where $y$ is the extra dimension. 
When $\Ly$ is infinite, the system is ordered and the correlation length $\xi$ is infinite (Goldstone modes are massless). 
Now, if $\Ly$ is reduced and made finite, a new length scale $\xiSR(\Ly)$ appears in the system,
which we may call the \emph{symmetry restoration scale}.
For scales smaller than $\xiSR(\Ly)$, the system is ordered (symmetry broken), while for scales larger that $\xiSR(\Ly)$, the system is disordered by the fluctuations of the massive Goldstone modes and the symmetry is restored.

For $\Ly$ finite, $\xiSR(\Ly)$ must be finite. This is because, if it were infinite, then for any finite $\Ly$,
the system would be ordered along the $y$ direction, and would therefore effectively become (1+1)-dimensional.
But this would imply that the (1+1)-dimensional system is ordered and has a broken continuous symmetry, 
which is impossible due to the Coleman--Mermin--Wagner theorem.
In other words, as $\Ly$ is reduced from being infinite to finite, the Goldstone modes pick up a mass $\sim \xiSR^{-1}$, which determines the scale at which the system disorders.

Let us assume that we can make $\Ly$ small enough such that $\xiSR(\Ly) \gg \Ly$. 
Again, the physics is effectively frozen in the $y$ direction and it can be explicitly integrated over in the low-energy effective action to obtain
\begin{align}
  S_{0} &= \frac{\Ly}{2g^2} \int dx\, dt\,  \Tr (\del_{\mu} P )^2 \nonumber
\end{align}
We see that the (1+1)-dimensional theory has an effective coupling $\geff^2 = g^2/\Ly$.  
Now, we can use knowledge of the (1+1)-dimensional theory.
Asymptotic freedom in the (1+1)-dimensional Grassmannian models implies that the system develops an exponentially large length scale $\xiSR \propto e^{\Ly/(\beta_0 g^2)} $, where $\beta_0$ is the leading $\beta$ function coefficient. Therefore, for small enough $\Ly$, we get $\xiSR(\Ly) \gg \Ly$, validating the dimensional reduction scenario.

In the above argument, we assumed that $\xiSR(\Ly) \gg \Ly$ by starting from $\Ly$ large.
In principle, it possible that this is not valid for some regime of $\Ly$ and dimensional reduction does not occur.
However, an independent argument can be given by considering $\Ly=\Order(1)$ very small. A nonlinear sigma model analysis of the spin ladders with finite $\Ly$ shows that $\xiSR$ is finite (in lattice units) and grows exponentially in $\Ly$, and therefore we can always ensure that $\xiSR(\Ly) \gg \Ly$ \cite{sierra_nonlinear_1996}.

We note that the above discussion is valid for any $\theta$, since the \ac{UV} physics of asymptotic freedom is the same in all cases.  For $\theta=0$, there is only one length scale in the system, and therefore the symmetry restoration length scale $\xiSR$ and the correlation length $\xi$ coincide.  However, for $\theta\neq0$, $\xiSR$ and $\xi$ are different, due to nonperturbative effects from the $\theta$ term.
$\xiSR$ sets the \ac{UV} scale of asymptotic freedom, while $\xi$ sets the scale for \ac{IR} physics.
In general, we have the following hierarchy of length scales
\begin{align}
  \xi(\theta) \gtrsim \xiSR \gg \Ly \gg a
\end{align}
where $a$ is the lattice spacing.
Indeed, for the $\theta=\pi$ $\CP(1)$ theory, the nonperturbative effects are strong enough that the system becomes gapless with $\xi = \infty$.

\emph{Choice of representations.}
We emphasize this approach strongly depends the representations at each site.
While in principle, for a given $\gr$, we can choose any $\Nc$ to get a lattice model, not every such model will have a continuum limit in the above prescription. For the alternating configuration, 
the representations $R, \bar R$ needs to be sufficiently large.
This is because for the D-theory prescription to work, we need the two-dimensional ($\Lx, \Ly \to \infty$) theory to be in massless symmetry-broken phase.  Read and Sachdev \cite{read_largen_1991, read_features_1989, read_valencebond_1989} studied the two-dimensional model at large-$N$ and large-$\Nc$ limits, and obtained the result that the system is in the N\'eel phase for $\Nc \geq \kappa N$, where $\kappa$ is a constant of order 1.
These semiclassical arguments suggest that as long as we choose $n \geq \kappa N$, we are guaranteed the right continuum limit as $\Ly \to \infty$.
For example, for $k=1$, it has been numerically shown that it suffices to choose the fundamental representation for $N=2,3,4$, while larger $\Nc$ representations are needed to obtain a N\'eel ordered ground state in two spatial dimensions for $N \geq 5$ \cite{harada_eel_2003}.


\section{Lattice formulations of the Grassmannian models with Anomalies}
\label{sec:luscher}

In this section, we discuss lattice regularizations of the $\Gr_k(N)$ models in which both $\Csym$ and $\PSU(N)$ act onsite. As mentioned above, this can only be compatible with the anomaly of the continuum theory if the anomaly is present on the lattice.

In the continuum, the charge conjugation symmetry at $\theta=\pi$ is not manifest in the sense that the action is not invariant but is shifted by $2\pi i Q$. It is thanks to the integer quantization of $Q$ that the integrand $\exp(-S)$ of the path integral is preserved by $\Csym$.

The statement of the mixed anomaly between the $\PSU(N)$ symmetry and $\Csym$ is that when background gauge fields for the former are present, the partition function is no longer invariant under the latter \cite{Gaiotto:2017yup, Komargodski:2017dmc, komargodski_walls_2018, cordova_anomalies_2020, Dunne:2018hog}. At a technical level, this is due to the fact that the presence of such background fields replaces the topological charge quantization in units of $1$, which as we just said was necessary for $\Csym$ invariance at $\theta=\pi$, by quantization in units of $1/N$. This anomaly can therefore be regarded as a consequence of the more basic fact that in the presence of background fields for the $\PSU(N)$ symmetry, the period of the $\theta$ dependence should become $2\pi N$ so that the partition function fails to be $2\pi$ periodic in $\theta$ once the $\PSU(N)$ symmetry is gauged. We may describe this more basic situation as a mixed anomaly between $\PSU(N)$ symmetry and $2\pi$ periodicity in $\theta$.

All this is equally true in the lattice models we shall describe below. 

\subsection{Berg--L\"uscher $\theta$ term}

We begin with a discretization on a two-dimensional triangular Euclidean spacetime lattice based on the continuum action \eqref{eq:S-nlsm} \cite{divecchia_lattice_1981}:
\begin{align}
  S\lbrack P\rbrack = -\frac{1}{g^2} \tsum_{\< x y\> } \Tr P_x P_y - i \theta \tsum_{\plaq} q_\plaq,
  \label{eq:CP_lattice_action}
\end{align}
where the first sum runs over all nearest-neighbor bonds $\<xy\>$ on the triangular lattice, the second sum runs over all positively oriented triangular plaquettes $p$. The fields 
$P_x$ are Hermitian $N \times N$ projection matrices such that $\Tr P_x = k$, and $q_\plaq$ is the topological charge density to be defined below. 
The choice of a triangular lattice is motivated by the definition of the topological charge density. 
At $g \to 0$ with a fixed $\theta$, the lattice action in \cref{eq:CP_lattice_action} defines a continuum limit of the $\gr$ \ac{NLSM}.
For $\theta = 0$, this model has been extensively studied for various $N$.
There is a symmetry of the action under
$P_x \mapsto v P_x v^\dagger$, for $v$ an element of $\SU(N)$.  Note that since the $P_x$ do not transform under the center of $\SU(N)$, the correct (faithful) symmetry group of the action is $\SU(N)/\ZZ_N \cong \PSU(N)$. Charge conjugation acts by sending $P_x \mapsto P_x{}^{\!*}$, and so clearly preserves the action at $\theta=0$.

For $\Gr_1(N) = \CP(N-1)$, Berg and L\"uscher \cite{berg_definition_1981} have defined $4\pi q_\plaq$ to be the minimal signed area spanned by the geodesic triangle in the $\CP(N-1)$ manifold with vertices $P_x, P_y, P_z$, where $x,y,z$ are the vertices of the plaquette $\plaq$ in counterclockwise order. Explicitly,
\begin{align}
      \exp(2\pi i q_\plaq)
      = \frac{\Tr P_x P_y P_z}{\lvert \Tr P_x P_y P_z \rvert}.
      \label{eq:lattice_CP_top_charge}
\end{align}
The branch ambiguity implied by this equation is fixed by requiring that $q_p$ lie in the interval $(-\frac{1}{2},\frac{1}{2})$. (The same should be understood in similar formulas below.) 
The total topological charge $Q = \sum_{p} q_\plaq$ is obtained by summing over all positively oriented plaquettes. By construction it is an integer for periodic boundary conditions (see \cref{sec:theta}). 
Notice that $\Csym \colon q_p \mapsto -q_p$. Therefore, the $\Csym$ symmetry at $\theta=\pi$ here is not manifest, just as in the continuum.

We shall generalize this definition of the topological charge for $\Gr_k(N)$ fields below. But let us first demonstrate the existence of the anomaly in $\CP(N-1)$ model.

\begin{figure}
  \centering
  \includegraphics{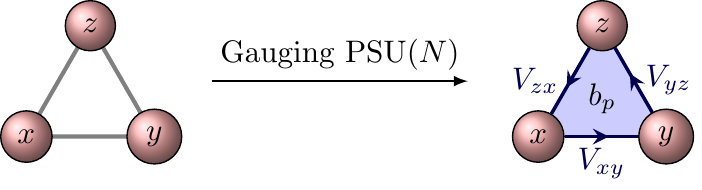}
  \caption{
    The Berg--L\"uscher $\theta$ term is defined on a triangulated lattice such that the $\gr$ field lives on the nodes $x, y, z$.  The $\theta$ term is defined on the triangular plaquette $p = \< x y z \> $. To gauge the $\PSU(N)$ symmetry, we activate a $\SU(N)$ link field $V_\ell$ as well as a $\ZZ_N$ plaquette field $b_p$.}
  \label{fig:berg-luscher}
\end{figure}

Our strategy is to background-gauge the $\PSU(N)$ symmetry (see \cref{fig:berg-luscher}), and then examine the $\theta$ dependence of the partition function. We start by introducing an external link field $V_\ell \in \SU(N)$. The action should incorporate it in such a way that it be invariant under the local $\SU(N)$ transformations
\begin{align}
  P_x  \mapsto v_x  P_x  v_x{}^{\!\dag}, \ 
  V_{xy}  \mapsto v_x  V_{xy}  v_y{}^{\!\dag}.
  \label{eq:0form_GT}
\end{align}
This can be achieved by modifying the kinetic part of the action in \cref{eq:CP_lattice_action} to
\begin{align}
  \Tilde{S}_0\lbrack P, V \rbrack \equiv -\beta \tsum_{\<xy\>} \Tr P_x V_{xy} P_y V_{yx}.
  \label{eq:gauged_lattice_zinetic_term}
\end{align}
and the topological charge density in \cref{eq:lattice_CP_top_charge} to 
\begin{align}
    \exp(2\pi i \hat{q}_p)
    \equiv 
    \frac{\Tr P_x V_{xy} P_y V_{yz} P_z V_{zx}}{\lvert \Tr P_x V_{xy} P_y V_{yz} P_z V_{zx} \rvert}.
    \label{eq:qhat}
\end{align}
We note that $\hat{Q}=\sum \hat{q}_\plaq$ is still an integer (see Appendix \ref{sec:theta}). 

However, we are gauging not $\SU(N)$ but $\PSU(N)$. Two elements of the former that differ from each other only by an $N$th root of unity are identified as elements of the latter. In view of this fact, we postulate a gauge invariance under local $\ZZ_N$ transformations of the link fields,
\begin{align}
    V_\ell \mapsto V_\ell 
    \exp(2 \pi i n_\ell / N)
    \label{eq:1form_GT_prelim}
\end{align}
with $n_\ell \in \ZZ_N$. 
The new kinetic term is already invariant under this gauge transformation. The topological term, however, is not; it transforms as
\begin{align}
    \exp(2\pi i \hat{q}_\plaq)
    \mapsto \exp(2\pi i \hat{q}_\plaq)
    \tprod_{\ell \in \plaq} \exp(2\pi i n_\ell/N).
\end{align}
To fix this, we further introduce an external plaquette field $b_\plaq \in \ZZ_N$ with a gauge transformation rule chosen to precisely compensate for the noninvariance of $\hat{q}_\plaq$. Thus, if instead of \eqref{eq:1form_GT_prelim} we postulate a gauge invariance under the combined transformations
\begin{align}
    V_\ell &\mapsto V_\ell 
    \exp(2 \pi i n_\ell /N)
    , \nonumber\\
    b_\plaq &\mapsto b_\plaq + \tsum_{\ell \in \plaq} n_\ell,
    \label{eq:1form_GT}
\end{align}
then the topological charge density can be made invariant by modifying it to 
\begin{align}
    \exp ( 2\pi i \Tilde{q}_\plaq ) 
    \equiv \exp ( 2\pi i \hat{q}_\plaq ) 
    \exp( - 2 \pi i b_\plaq/N)
\end{align}

Thus, with the action 
$\Tilde{S} = \Tilde{S}_0 - i \theta \tsum \Tilde{q}_\plaq$
we have succeeded in coupling the theory to background fields for the $\PSU(N)$ symmetry in a totally gauge invariant manner. 
The question now is whether the partition function
\begin{align}
  \Tilde{Z}\lbrack V,b,\theta \rbrack = \int \mathcal{D}P\, \exp( -\Tilde{S}_0\lbrack P, V \rbrack + i \theta \Tilde{Q}\lbrack P, V, b \rbrack)
\end{align}
is still $2\pi$ periodic in $\theta$.

The answer is no. 
This is because in the presence of the $\PSU(N)$ background gauge field, the gauge-invariant topological charge $\Tilde{Q} = \sum_\plaq \Tilde{q}_\plaq$ is in general quantized not in units of $1$ but in units of $1/N$, so that the period of the $\theta$ dependence is in general enlarged to $2\pi N$. More precisely, we have 
\begin{align}
    \exp(2\pi i \Tilde{Q}) 
    &=  \exp(2\pi i \hat{Q}) \exp(-2\pi i B/N)
    \nonumber\\
    &= 
    \exp(-2\pi i B/N),
\end{align}
(we have written $B \equiv \sum_\plaq b_p$) 
and this phase factor, which for arbitrary $b_\plaq$ can be any $N$th root of unity, is precisely the factor by which the partition function is changed upon a shift of $\theta$ by $2\pi$:
\begin{align}
    \frac{\Tilde{Z}\lbrack V,b,\theta+2\pi\rbrack}{\Tilde{Z}\lbrack V,b,\theta\rbrack}
    =
    \exp(-2\pi i B/N).
\end{align}
As already mentioned, the \Csym{} invariance at $\theta=\pi$ depends crucially on the topological charge being an integer, and so gauging $\PSU(N)$, which fractionalizes the topological charge, explicitly breaks \Csym{} invariance at $\theta=\pi$. 

\renewcommand\Id{\mathbf{1}}

A nearly verbatim discussion applies to the Grassmannian model, to which we now return.
However, the topological charge is not simply given by the Berg--L\"uscher formula \eqref{eq:lattice_CP_top_charge}.
In this case, it is convenient to write the projector valued field as $P_x = \phi_x \phi_x{}^{\!\dagger}$, where $\phi_{x}$ are $N\times k$ complex matrices, such that $\phi_x{}^{\!\dagger} \phi_x = \Id_k $.  The $k$ columns of $\phi$ specify a $k$-dimensional subspace onto which $P_x$ projects.
In this notation, 
the natural generalization of the Berg--L\"uscher $\theta$ term to this case is given by
\begin{align}
    \exp(2\pi i q_\plaq)
    = \tprod_{\<xy\> \in p}
    \frac{\det \grfield_x{}^{\!\dag} \grfield_y}{\lvert \det\grfield_x{}^{\!\dag} \grfield_y \rvert}.
    \label{eq:gr-top-charge}
\end{align}
(See \cref{sec:theta} for a derivation.)
Note that for $k=1$, the determinant is the same as the trace, and therefore this reduces to \cref{eq:lattice_CP_top_charge} for $\CP(N-1)$.

With the correct definition of the topological charge in hand, we can now detect the mixed anomaly between $2\pi$ periodicity in $\theta$ and $\PSU(N)$ symmetry as before. We introduce an $\SU(N)$ link field $V_\ell$ and a $\ZZ_N$ plaquette field $b_\plaq$ and modify the kinetic term to \eqref{eq:gauged_lattice_zinetic_term}, and the topological charge density to
\begin{align}
    \exp ( 2\pi i \Tilde{q}_\plaq ) 
    \equiv \exp ( 2\pi i \hat{q}_\plaq ) 
    \exp( - 2\pi i k b_\plaq/N),
    \label{eq:gr-top-charge-gauged}
\end{align}
where
\begin{align}
    \exp(2\pi i \hat{q}_\plaq)
    \equiv 
    \tprod_{\<xy\> \in p}
    \frac{\det \grfield_x{}^{\!\dag} V_{xy}\grfield_y}{\lvert \det\grfield_x{}^{\!\dag} V_{xy} \grfield_y \rvert}
\end{align}
We note that in the last exponent of \eqref{eq:gr-top-charge-gauged}, the factor of $k$ appears because of the determinant of $k\times k$ matrices in \eqref{eq:gr-top-charge}.
After literal repetition of the preceding argument, we find
\begin{align}
    \frac{\Tilde{Z}\lbrack V,b,\theta+2\pi\rbrack}{\Tilde{Z}\lbrack V,b,\theta\rbrack}
    =
    \exp( - 2\pi i k B/N).
    \label{eq:theta-anomaly-lattice}
\end{align}
We note that the anomalous phase for the $\gr$ models is $k$ times that of the $\CP(N-1)$ model. 

\subsection{Lattice gauge theory formulation}

We have given one lattice formulation of the two-dimensional $\Gr_k(N)$ model that explicitly exhibits the anomaly, but we wish to emphasize that the anomaly was inevitable given that the symmetries were realized onsite. Thus, in this subsection, we present a different lattice regularization in which the same is true. It is based on the continuum action
\begin{align}
    S 
    = 
    \int d^2 x\,  \lvert \del_\mu \grfield - \grfield A_\mu\rvert^2 
    + \frac{\theta}{2\pi} \int \Tr d A,
\end{align}
where $\grfield$ is a $N\times k$ complex matrix--valued scalar field satisfying $\grfield^\dag \grfield = \mathbf{1}_k$, $A_\mu$ is a $\U(k)$ gauge field and we use the notation $\lvert M \rvert^2 \equiv \Tr M^\dag M$ for matrices $M$. (The gauge field is auxiliary and can be trivially integrated out to write the action entirely in terms of $\grfield$.) The relation to the projector-valued field $P$ is simply $P=\grfield \grfield^\dag$. 

The standard discretization\footnote{For $\CP(N-1)$ an alternative discretization via the modified Villain formalism was given in Ref.~\cite{Sulejmanpasic:2019ytl}, which can be shown to also explicitly exhibit the anomaly.} on a square lattice gives
\begin{align}
    S &= 
    -\beta \tsum_{\langle xy \rangle} \Tr \grfield_x{}^{\!\dag} \grfield_y U_{yx} - i \theta \tsum_p q_p,
\end{align}
where the $\grfield_x$ are complex $N\times k$ matrices defined on sites such that $\grfield_x{}^{\!\dag} \grfield_x = \mathbf{1}_k$, the $U_\ell$ are elements of $\U(k)$ defined on links, and the topological charge $q_p$ in a plaquette $p$ is defined by
\begin{align}
    \exp(2\pi i q_p) \equiv \tprod_{\ell \in p} \det U_\ell
\end{align}
By taking the product of this last equation over all plaquettes, we clearly see that $Q=\sum q_p$ is an integer. Moreover, it is clear that $q_p$ has the right continuum limit, using the correspondence $\det U_\mu(x) \to \exp(a \Tr A_\mu(x))$ ($a$ is the lattice spacing). For $\CP(N-1)$, this reduces to the definition used in \cite{PhysRevLett.53.637}. 
The lattice action has a gauge symmetry under
\begin{align}
    \grfield_x \mapsto \grfield_x u_x, \ U_{xy} \mapsto u_x{}^{\!\dag} U_{xy} u_y,
\end{align}
where $u_x \in \U(k)$. It also has a global symmetry under
\begin{align}
    \grfield_x \mapsto v \grfield_x, \ 
    U_\ell \mapsto U_\ell,
\end{align} 
for $v \in \SU(N)$. But for $v$ in the center of $\SU(N)$, i.e., $v = \exp(2\pi i n/N) \mathbf{1}_N$, this transformation is equivalent to a gauge transformation with $u_x = \exp(2\pi i n/N) \mathbf{1}_k$, so as before the true symmetry group is $\PSU(N)$. Finally, we have charge conjugation, which acts as
\begin{align}
    \Csym \colon \grfield_x \mapsto \grfield_x{}^{\!*}, \ 
    U_\ell \mapsto U_\ell{}^{\!*}.
\end{align}
In particular, we have $\Csym \colon q_p \mapsto -q_p$ so $\Csym$ invariance is manifest at $\theta=0$ but not at $\theta=\pi$, as before.

The demonstration of the anomaly here more or less follows the same lines as the above. We introduce an external gauge field for the $\PSU(N)$ symmetry via a pair consisting of a link-field $V_\ell \in \SU(N)$ and a plaquette-field $b_p\in \ZZ_N$
with the gauge transformation rule
\begin{align}
    V_{xy} &\mapsto v_x V_{xy} v_y{}^{\!\dag} \exp( 2\pi i n_{xy}/N), \nonumber \\ 
    b_p &\mapsto b_p + \tsum_{\ell \in p} n_\ell,
    \label{eq:background-gt}
\end{align}
where $v_x \in \SU(N)$ and $n_\ell \in \ZZ_N$. 
These background fields may be coupled to the dynamical fields in a gauge-invariant fashion as follows. We replace the kinetic term by
\begin{align}
    \tsum_{\<xy\>} \Tr \grfield_x{}^{\!\dag} V_{xy} \grfield_y U_{yx},
\end{align}
which is invariant under the transformation \eqref{eq:background-gt} of the background fields combined with the following transformation of the dynamical fields:
\begin{align}
    \grfield_x \mapsto v_x \grfield_x, \ 
    U_\ell \mapsto U_\ell \exp(2\pi i n_\ell/N).
\end{align}
The nontrivial transformation rule for $U_\ell$ then forces us to replace the topological charge density by
\begin{align}
    \exp(2\pi i \Tilde{q}_p) 
    \equiv \exp(-2\pi i k b_p/N) \tprod_{\ell \in p} \det U_\ell.
\end{align}
Taking the product of the last equation over all plaquettes, we get
\begin{align}
    \exp(2\pi i \Tilde{Q}) = \exp(-2\pi i k B/N),
\end{align}
from which the same anomaly \eqref{eq:theta-anomaly-lattice} follows.



\section{Conclusion}
\label{sec:conclusion}

Anomalies have long been a powerful way to probe nonperturbative aspects of \acp{QFT}. 
For lattice regularizations, they imply constraints on the way anomalous symmetries may be realized. 
Indeed, a \ac{QFT} with an anomalous symmetry $G$ is such that $G$ cannot be gauged; the same must be true for any symmetric lattice regulator thereof.

In this work, we have explored the connection between discrete anomalies and lattice regularizations for a class of \acp{NLSM} with target space given by the Grassmannian manifold $\gr$, which includes the well-known $\CP(N-1)$ models as the case with $k=1$.
These models have a mixed anomaly between
$\PSU(N)$ and $\Csym$ symmetries. 
\cite{Dunne:2018hog, Gaiotto:2017yup, Komargodski:2017dmc, komargodski_walls_2018, cordova_anomalies_2020}.

The question we have tried to address in this work is: What do these anomalies imply for constructing a symmetric lattice regulator? 
Generally, it is thought that an anomaly in a symmetry implies that on the lattice the symmetry cannot be realized onsite, because then one could always gauge the symmetry by introducing appropriate link variables. However, this is incorrect -- there is also the possibility that the anomaly is \emph{explicitly} reproduced on the lattice. 
This can be surprising, since it is often remarked that ``there are no anomalies on the lattice.''  
Indeed, one of the most well-known early constructions of the $\theta$ term in the (1+1)-dimensional $\CP(N-1)$ \ac{NLSM} due to Berg and L\"uscher \cite{berg_definition_1981}, already exhibits this mixed anomaly between $\PSU(N)$ and $\Csym$, as we have demonstrated in this work.

We also noted that in some cases, strictly speaking, there is no anomaly but a more subtle scenario referred to as a ``global inconsistency'' \cite{kikuchi_global_2017,Gaiotto:2017yup,tanizaki_vacuum_2017}.  In such cases, we argued that the usual no-go theorem gets modified and presents a new kind of obstruction, 
which prevents us from finding lattice regulators with manifest onsite symmetry for the two points of global inconsistency that can be continuously connected to each other.

Of course, it is possible that the anomalous symmetry is realized offsite, in which case it is not possible to gauge it in the usual way. This is another way that an anomaly may be reflected in a lattice regularization.
We demonstrated this possibility by constructing a new regularization for the $\gr$ \ac{NLSM} with a $\theta$ term, using the D-theory formulation. This regularization uses $\SU(N)$ spins as fundamental degrees of freedom with local antiferromagnetic interactions on a two-dimensional spatial $\Lx\times \Ly$ lattice. In the large-representation limit, the low-energy limit can be shown to be that of (1+1)--dimensional $\gr$ \ac{NLSM} for a fixed $\Ly$.  Extending on recent work for the $\O(3)$ \ac{NLSM} \cite{caspar_asymptotic_2022}, we argued that a similar construction allows us to obtain the obtain continuum limit of the $\gr$ \ac{NLSM} at \emph{arbitrary} non-zero $\theta$, by keeping $\gamma\Ly$ fixed as we take the $\Ly\to \infty$ limit.

For the $\CP(N-1)$ models, this provided the first completely sign-problem free lattice regularization of the $\theta$ vacua \cite{beard_efficient_2006,caspar_asymptotic_2022}.
In the case of $\SU(N)$ antiferromagnets in the $(k,\Nc)$ representation with $k\neq 1$, efficient sign-problem free algorithms are not known. Even though a direct numerical confirmation of our regularization is currently limited due to this, it is interesting to note that the sign-problem in $\SU(N)$ $(\Nr,\Nc)$ antiferromagnets in general is of a totally different character than the conventional Berg--L\"uscher $\theta$ term.
After all, $(\Nr, \Nc)$ $\SU(N)$ antiferromagnets have a sign problem even at $\theta=0$. Changing $\theta$ by adding staggered couplings does not make it more severe. This suggests that if an efficient algorithm for $\SU(N)$ antiferromagnets with $k\neq1$ representations can be found at $\theta=0$, we would also automatically have a sign-problem free approach to arbitrary nontrivial $\theta$ for the entire class of $\gr$ \ac{NLSM}. 

The discussion of anomalies and lattice regularizations presented in this work is especially relevant in the context of four-dimensional Yang--Mills theories.
Indeed, at $\theta=\pi$, pure $\SU(N)$ Yang--Mills theory exhibits a mixed anomaly involving time-reversal and center symmetry \cite{Gaiotto:2017yup} that is quite analogous to the anomaly studied here.
There already exist proposals for qubit regularizations for Yang--Mills as quantum link models \cite{brower_qcd_1999, brower_dtheory_2004}, while the conventional lattice regularization admits a topological construction for the $\theta$ term \cite{luscher_topology_1982}.
There seems to be a parallel discussion to be had in these cases. We leave this for a future publication.


\section*{Acknowledgments}

HS would especially like to thank Stephan Caspar for very valuable discussions and collaboration.
We would also like to thank Shailesh Chandrasekharan, David Kaplan, Hanqing Liu, Tin Sulejmanpasic, Yuya Tanizaki, Mithat \"Unsal, and Yi-Zhuang You for useful conversations, and special thanks to Roxanne Springer for initiating this collaboration.
HS was funded 
in part by the DOE QuantISED program through the theory consortium ``Intersections of QIS and Theoretical Particle Physics'' at Fermilab with Fermilab Subcontract No. 666484,
in part by the Institute for Nuclear Theory with US Department of Energy Grant No. DE-FG02-00ER41132,
and in part by U.S. Department of Energy, Office of Science, Office of Nuclear Physics, InQubator for Quantum Simulation (IQuS) under Award No. DOE (NP) DE-SC0020970.

\bibliography{refs}

\appendix

\section{Parametrizations of the Grassmannian manifold}
\label{sec:grasmannian}

The Grassmannian $\Gr_k(N)$ is the manifold parametrizing $k$-dimensional subspaces of a $N$-dimensional complex vector space.
Operationally, it can be defined as follows.
On an $N$-dimensional complex vector space, we pick a reference $k$-dimensional hyperplane $K_0$.
Let the hyperplane $K_0$ be represented by $k$-orthonormal complex vectors $K_0 = \lbrack \vec z_1, \dotsc, \vec z_k\rbrack $.
We may then denote the $k$-dimensional plane with the projector matrix
\begin{align}
  P_0 =  | \vec z_1 \>\< \vec z_1 | + \cdots + |\vec z_k \> \< 
\vec z_k |.
  \label{eq:proj-1}
\end{align}
Now, we may get any other $k$-dimensional hyperplane by performing an arbitrary $\U(N)$ rotation,
\begin{align}
  P_0 \mapsto P = U P_0 U^{\dagger} = \sum_{i=1}^{k} | U z_{i} \> \< U z_{i} |.
  \label{eq:proj-U(N)}
\end{align}
However, note that any unitary performed in the subspace $K^\perp_0$ orthogonal to $K_0$ will not yield a new hyperplane.
That's $\U(N-k)$ worth of unitaries which leave $P_0$ invariant.
Morever, any unitary in the subspace $K_0$ will only change our choice of reference vectors $\vec z_1, \dotsc, \vec z_k$, but will not change the hyperplane they define. That's another $\U(k)$ worth of unitaries.
We say that $\U(k) \times \U(N-k)$ is the isotopy group (or the little group) of the reference vector $P_0$.
The elements of $\U(N)$ that actually generate a new hyperplane are
\begin{align}
  \frac{\U(N)}{\U(k) \times \U(N-k)},
\end{align}
which is the precisely Grassmanian manifold $\Gr_k(N)$.

We can easily see from \cref{eq:proj-1,eq:proj-U(N)} that $P$ satisfies $P^2 = P = P^\dagger$, and $\Tr P = k$.
Therefore, the Grasmannian manifold can also be defined as the set of $N\times N$ complex matrices such that
\begin{align}
  \Gr_k(N) = \left\{P \in M_N(\CC) :  P^2 = P, \Tr P = k \right\}.
\end{align}
We note that Read and Sachdev \cite{read_features_1989} use the parametrization in terms of $Q$ variables, where $P = \frac12(1 + Q)$, such that $Q$ is a Hermitian $N \times N$ matrix with $Q^2=1$ and $\Tr Q = N - 2 k$.

We can also represent the $k$ vectors $\vec z_i$ as the columns of a $N \times k$ complex matrix $\phi$ matrix such that $\phi^\dagger \phi = \Id_{k}$, which allows us to rewrite \cref{eq:proj-1} as $P = \phi \phi^\dagger $.  Note that if the $\gr$ field theory is written in terms of the $\phi$ (or $\vec z_i$) fields, we must add a $\U(k)$ gauge field to account for the redundancy in the choice of reference vectors $\vec z_i$:
\begin{align}
  Z = \int \mathcal{D}P\, e^{-S\lbrack P\rbrack }
  = \int \mathcal{D} \phi\, \mathcal{D} \phi^\dagger\, \mathcal{D} A \,\, e^{-S'\lbrack \phi, \phi^\dagger, A\rbrack }.
\end{align}

It may also be useful to point out that the conventional formulation of the $\Gr_{1}(2)=\CP(1)$  or the $O(3)$ model in terms a real unit vector $\vec n$ (as in \cref{eq:S-O3}) is related to the projector formalism by
\begin{align}
  P = \tfrac12 (\Id_{2} + \vec n \cdot \vec \sigma), \quad \text{or}\quad \vec n = \Tr(P \vec \sigma).
\end{align}
Note that charge conjugation $P \mapsto P^{*}$ sends
\begin{align}
  (n_1, n_2, n_3) \mapsto (n_1, -n_2, n_3),
\end{align}
which differs from the given in \cref{eq:C-O(3)} $(\vec n \mapsto - \vec n)$ by an additional $\pi$-rotation about the second axis.


\section{Some properties of the Berg--L\"uscher lattice $\theta$ term}
\label{sec:theta}

In this section, we discuss some properties of the Berg--L\"uscher lattice $\theta$ term and provide its generalization for $\gr$ fields. 

It will be convenient to use the redundant parametrization of $\gr$ by $N\times k$ complex matrices $\phi$ satisfying $\phi^\dag \phi = \mathbf{1}_k$. 
In this notation, the Berg--L\"uscher $\theta$ term for $\CP(N-1) = \Gr_1(N)$ reads
\begin{align}
    \exp(2\pi i q_p) = \tprod_{\<xy\> \in p} \frac{(\phi_x,\phi_y)}{\lvert(\phi_x,\phi_y)\rvert},
    \label{eq:berg-luscher-alt}
\end{align}
where in this section we use the notation $(.\,,.)$ for the Hermitian scalar product of vectors. We see that $\exp(2\pi i q_p)$ takes the form of a product of link variables in $\U(1)$ around the boundary of $p$. Thus, in taking the product of this equation over all positively oriented plaquettes, the link variables cancel in pairs, so that
\begin{align}
    \exp(2\pi i Q) = \tprod_p \exp(2\pi i q_p) = 1.
\end{align}
Hence, the total topological charge $Q$ is an integer. 

We can now easily address an issue which was elided in \cref{sec:luscher}. There, background $\SU(N)$ link variables $V_\ell$ were introduced, and the topological charge $q_p$ got replaced by $\hat{q}_p$ which depends on the $V_\ell$ as in \eqref{eq:qhat}. An important property of $\hat{q}_p$ (which was not explained in the main text) is that $\hat{Q} = \sum_p \hat{q}_p$ be integer-quantized. We can now clearly see that this is so, as we can rewrite \eqref{eq:qhat} in the new notation as
\begin{align}
    \exp(2\pi i \hat{q}_p) = \tprod_{\<xy\> \in p} \frac{(\phi_x, V_{xy} \phi_y)}{\lvert(\phi_x, V_{xy} \phi_y)\rvert}. 
\end{align}

Let us now find the generalization of \eqref{eq:berg-luscher-alt} for the $\gr$ model. We first note that it is possible to realize $\gr$ as a submanifold of the complex projective space $P(\wedge^k \CC^N) \cong \CP \big(\binom{N}{k}-1 \big)$ via the Pl\"ucker embedding, which is given by
\begin{align}
    \Gr_k(N) &\to P(\wedge^k \mathbb{C}^N) \nonumber \\
    \lbrack z_1,\ldots,z_k \rbrack &\mapsto \lbrack z_1 \wedge \cdots \wedge z_k \rbrack
\end{align}
Here $z_1,\ldots,z_k$ is a set of orthonormal column vectors in $\CC^N$, so $\lbrack z_1,\ldots,z_k \rbrack$ denotes the $k$-dimensional subspace of $\CC^N$ generated by these vectors, and $\lbrack z_1 \wedge \cdots \wedge z_k \rbrack$ denotes the $1$-dimensional subspace in $\wedge^k \CC^N$ generated by $z_1 \wedge \cdots \wedge z_k$. 

This embedding is helpful here because it allows us to define the topological charge of a $\Gr_k(N)$ field by mapping it to a $P(\wedge^k \CC^N)$ field and then using the definition of the topological charge for $P(\wedge^k \CC^N)$ fields.
To see the validity of this claim, consider a $P(\wedge^k \CC^N)$ field $\Phi$ of the form $\Phi = z_1 \wedge \cdots \wedge z_k$ with $(z_i,z_j) =\delta_{ij}$, which by the Pl\"ucker embedding corresponds to a $\Gr_k(N)$ field $\phi = (z_1,\ldots,z_k)$. Then the topological charge $Q$ of $\Phi$ as a $P(\wedge^k \CC^N)$ field is given by
\begin{align}
    2 \pi Q &= \int d\big(\Phi,d\Phi \big) \nonumber\\
    &= \int d\big(z_1 \wedge \cdots \wedge z_k, d\{z_1 \wedge \cdots \wedge z_k\} \big) \nonumber\\
    &= \int d\tsum_{j=1}^k\big(z_1 \wedge \ldots \wedge z_k, z_1 \wedge \cdots \wedge dz_j \wedge \ldots \wedge z_k \big) \nonumber\\
    &= \int d\tsum_{j=1}^k \big( z_j, dz_j) \nonumber\\
    &= \int d \Tr \grfield^\dag d \grfield.
\end{align}
In the fourth line, we have used the formula for the scalar product in $\wedge^k \CC^N$ induced by the standard scalar product in $\CC^N$:
\begin{align}
    (z_1 \wedge \cdots \wedge z_k , w_1 \wedge \cdots \wedge w_k) = \det_{ij} (z_i,w_j).
    \label{eq:ext-scalar-prod}
\end{align}
The last line is just the topological charge of $\grfield$ as a $\Gr_k(N)$ field. Our claim is thus proved. 

Back to the lattice, we see with the help of formula \eqref{eq:ext-scalar-prod} that the Berg--L\"uscher definition \eqref{eq:berg-luscher-alt} (replacing there $\CP(N-1)$ by $P(\wedge^k \CC^N)$ and $\phi_x$ by $\Phi_x$) gives the desired definition of the topological charge for lattice $\gr$ fields $\phi_x$ as
\begin{align}
    \exp(2\pi i q_\plaq)
    = \tprod_{\<xy\> \in p}
    \frac{\det \grfield_x{}^{\!\dag} \grfield_y}{\lvert \det\grfield_x{}^{\!\dag} \grfield_y \rvert}.
\end{align}
For completeness, we note that it is possible to give an expression in terms of the projector-valued field $P_x$:
\begin{align}
    \exp(2\pi i q_\plaq)
    = \frac{\Tr \wedge^k P_x \cdot \wedge^k P_y \cdot \wedge^k P_z}{\lvert \Tr \wedge^k P_x \cdot \wedge^k P_y \cdot \wedge^k P_z\rvert},
\end{align}
where $\wedge^k P_x$ is the operator on $\wedge^k \CC^N$ induced by the operator $P_x$ on $\CC^N$. 


\section{Continuum presentation of the $\gr$ model anomalies}
\label{sec:anomalies}

This appendix reviews the continuum presentation of the mixed anomaly in the two-dimensional $\Gr_k(N)$ model between $\PSU(N)$ symmetry and charge conjugation $\Csym$ \cite{Dunne:2018hog}, which the lattice presentation in the main text mirrors closely. The analysis is very similar to that of the two-dimensional $\CP(N-1)$ model \cite{Gaiotto:2017yup, Komargodski:2017dmc, komargodski_walls_2018, cordova_anomalies_2020}.

We shall work with the continuum action
\begin{align}
    S = \int \lvert d \phi - \phi a \rvert^2
    + \frac{\theta}{2\pi} \int \Tr da. 
\end{align}

To detect the mixed anomaly, we shall need to activate a background gauge field for the $\PSU(N)$ symmetry, which can be regarded as a pair of fields consisting of a $\U(N)$ gauge field $A$ and a $\U(1)$ 2-form gauge field $B$ that satisfy the constraint $N B = \Tr d A$ \cite{Kapustin:2014gua}. (In our conventions, $B$ is pure imaginary.) This pair transforms under a generalized gauge transformation of the form
\begin{align}
    A \mapsto gAg^{-1} + g dg^{-1} + \lambda \mathbf{1}_N, 
    \  
    B \mapsto B + d\lambda
    \label{eq:background-gt-cont}
\end{align}
where $g$ is a $\U(N)$-valued function and $\lambda$ is a $\U(1)$ 1-form gauge field. 
To couple these background fields to the dynamical fields in a gauge invariant fashion, we replace the action by
\begin{align}
    \tilde{S} = \int \lvert d \grfield  + A \grfield - \grfield a \rvert^2 
    + \frac{\theta}{2\pi} \int \Tr (da - B \mathbf{1}_k).
\end{align}
This is invariant under the transformation \eqref{eq:background-gt-cont} of the background fields provided we also make the following transformation on the dynamical fields:
\begin{align}
    \grfield \mapsto g \grfield , 
    \ 
    a \mapsto a + \lambda \mathbf{1}_k.
    \label{eq:dynamical-gt-cont}
\end{align}
It is now straightforward to find that the partition function in the presence of the background fields satisfies
\begin{align}
    \tilde{Z}\lbrack A,B,\theta+2\pi\rbrack 
    = \tilde{Z}\lbrack A,B,\theta\rbrack  \exp(k \tint B),
    \label{eq:theta-anomaly}
\end{align}
which indicates a mixed anomaly between the $\PSU(N)$ symmetry and the $2\pi$-periodicity in $\theta$. An important dynamical consequence of this result is that the theory cannot be trivially gapped for all values of $\theta \in \lbrack 0, 2\pi \rbrack$. 

To go further, we consider the charge conjugation symmetry at $\theta =0,\pi$. Charge conjugation acts by complex conjugating all fields:
\begin{align}
    \Csym\colon 
    \grfield \mapsto \grfield^*,\ 
    a \mapsto a^*, \ 
    A \mapsto A^*, \ 
    B \mapsto B^*.
\end{align}
But in the action, complex conjugating the fields is equivalent to reversing the sign of $\theta$; i.e.,
\begin{align}
    \tilde{S}\lbrack \grfield^*,a^*,A^*,B^*,\theta\rbrack 
    =
    \tilde{S}\lbrack \grfield,a,A,B,-\theta\rbrack 
\end{align}
Thus, at $\theta=0$, charge conjugation leaves the partition function invariant, while at $\theta=\pi$, it is equivalent to shifting $\theta$ by $-2\pi$, which by \eqref{eq:theta-anomaly} causes the partition function to pick up the phase factor $\exp(-k \int B)$:
\begin{align}
    \Csym\colon
    \begin{cases}
    \tilde{Z}\lbrack A,B,\theta=0\rbrack  \;\mapsto \tilde{Z}\lbrack A,B,\theta=0\rbrack,  \\
    \tilde{Z}\lbrack A,B,\theta=\pi\rbrack  \,\mapsto \tilde{Z}\lbrack A,B,\theta=\pi\rbrack  \exp(-k \tint B).
    \end{cases}
\end{align}
From this, we see that at $\theta=0$, there is no anomaly, while at $\theta=\pi$, there may or may not be an anomaly. The reason we cannot yet conclude that there is an anomaly at $\theta=\pi$ is that we have not yet considered the most general procedure of gauging $\PSU(N)$. Said differently, we are free to add to the action any gauge-invariant local ``counterterm'' depending only on the background fields. To have a genuine anomaly, it must be impossible to choose a counterterm that restores the symmetry. 

Suffice it to say, the only candidate for such a counterterm is of the form $p \int B$ with $p \in \ZZ_N$. If we therefore take for the partition function
\begin{align}
    \Tilde{Z}' \equiv \Tilde{Z} \exp(p\tint B),
\end{align}
then we shall have
\begin{align}
    \Csym\colon 
    \begin{cases}
    \Tilde{Z}'\lbrack A,B,0\rbrack \;\mapsto \,\Tilde{Z}'\lbrack A,B,0\rbrack \exp\{-2p \tint B\}, \\
    \Tilde{Z}'\lbrack A,B,\pi\rbrack \,\mapsto \,\Tilde{Z}'\lbrack A,B,\pi\rbrack  \exp\{- (2p+k)\int B \} .
    \end{cases}
\end{align}
Now, if there were a choice of $p$ such that \Csym{} invariance held at \emph{both} $\theta = 0$ and $\theta=\pi$, that is, if there were an integer $p$ such that $2p = 0$ mod $N$ and $2p + k = 0$ mod $N$, then we would have $k=0$ mod $N$, which is a contradiction ($1\leq k \leq N-1$). In fact, with a slightly more detailed argument \cite{Dunne:2018hog}, we find the following cases:
\begin{itemize}
    \item If $N$ is even and $k$ is odd, there is no $p$ that can restore \Csym{} invariance at $\theta = \pi$, so we have a genuine anomaly. To see this, note that if there were such a $p$ with $2p+k=0 $ mod $N$, then since $N$ is even, we would have $2p +k$ even. But this would then imply that $2p$ was odd, since $k$ is odd -- contradiction.
    \item Otherwise, it is possible to choose $p$ such that \Csym{} invariance is restored at $\theta = \pi$. Indeed, 
    \begin{itemize}
        \item for $N$ even and $k$ even, choose $p = N - \frac{1}{2}k$,
        \item for $N$ odd and $k$ even, choose $p = N - \frac{1}{2}k$,
        \item for $N$ odd and $k$ odd, choose $p = \frac{1}{2}(N-k)$.
    \end{itemize}
   But as we have already seen, we cannot repair \Csym{} invariance at $\theta = \pi$ without spoiling it at $\theta = 0$. In these cases, therefore, while we do not have a genuine anomaly, we do have a \emph{global inconsistency} \cite{kikuchi_global_2017,Gaiotto:2017yup,tanizaki_vacuum_2017}. 
\end{itemize}
From this, we can make the following inferences about the dynamics:
In the cases with a genuine anomaly, the theory cannot be trivially gapped at $\theta=\pi$.
In the cases with global inconsistency, there must exist some value of $\theta \in \lbrack 0,\pi\rbrack $ at which the theory is not trivially gapped. 


\section{The $\Gr_k(N)$ model from an $\SU(N)$ spin chain}
\label{sec:coherent-states}

\begin{figure}[tb]
  \centering
  \includegraphics{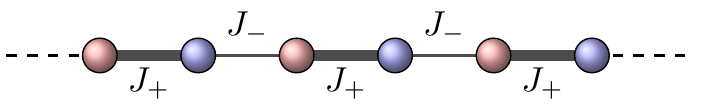}
  \caption{One-dimensional $\SU(N)$ Heisenberg antiferromagnetic chain with staggered couplings. The even and odd sites have conjugate representations of $\SU(N)$ as shown in \cref{fig:sublattices}, and the bonds have alternating strength $J_\pm = J (1 \pm \gamma)$.}
  \label{fig:1d-chain}
\end{figure}

In this appendix, we show that the two-dimensional $\Gr_k(N)$ model with a $\theta$ term arises as the large distance effective field theory of the antiferromagnetic $\SU(N)$ spin chain (see \cref{fig:1d-chain}) given by the Hamiltonian
\begin{align}
    H = J_-\tsum_n S_\alpha^\beta(n) \Sbar^\alpha_\beta(n)
    + J_+\tsum_n \Sbar^\alpha_\beta(n) S_\alpha^\beta(n+1).
    \label{eq:spin-chain-h}
\end{align}
where the $S_\alpha^\beta$ and $\Sbar^\alpha_\beta$ are the $\SU(N)$ spin operators in the $(k,p)$ and $(\overline{k,p})$ representations, respectively. (By the $(k,p)$ representation, we mean the representation corresponding to the rectangular Young tableau with $k$ rows and $p$ columns.) 

For $J_+ = J_-$, this was studied in Ref.~\cite{read_features_1989}, and in particular the result $\theta = \pi p$ was found. Our goal here is to show that by staggering the couplings, one should be able to achieve arbitrary values of $\theta$. 
This derivation generalizes a similar derivation for the $\CP(N-1)$ model in appendix B of Ref.~\cite{nguyen_winding_2022}.

\subsection*{Coherent states}

Introduce $N \times k$ harmonic oscillators
\begin{align}
    [a^{\alpha i},a^\dag_{j \beta}]=\delta_\beta^\alpha \delta_j^i,
\end{align}
where $\alpha,\beta = 1,\ldots,N$, $i,j=1,\ldots,k$, 
to represent the $\SU(N)$ spin operators by
\begin{align}
    S_\alpha^\beta = a^\dag_{i \alpha} a^{\beta i}.
\end{align}
The $(k,\Nc)$ representation of $\SU(N)$ is given by the subspace on which
\begin{align}
    N_i^j \equiv a^\dag_{i \alpha} a^{\alpha j} = \Nc \delta_i^j.
\end{align}
For any $N\times k$ complex matrix $\phi = (\phi^{\alpha i})$, the operator
\begin{align}
    \Phi &\equiv (k!)^{-\frac{1}{2}} a^\dag_{i_1 \alpha_1} \cdots a^\dag_{i_k \alpha_k} \epsilon^{i_1 \cdots i_k} \phi^{\alpha_1 1} \cdots \phi^{\alpha_k k} \nonumber \\
    &= (k!)^{-\frac{1}{2}} \det (a^\dagger \phi).
\end{align}
satisfies $[N_i^j,\Phi] = \delta_i^j$, so that $\Phi|0\rangle$ belongs to $(k,1)$. More generally, we have
\begin{align}
    [N_i^j,\Phi^\Nc] = \Nc \delta_i^j \Phi^\Nc,
\end{align}
so that $\Phi^\Nc |0\rangle$ is in the representation $(k,\Nc)$. 

We define coherent states for the representation $(k,\Nc)$ by
\begin{align}
    |\phi \rangle \equiv (\Nc !)^{-\frac{1}{2}} \Phi^\Nc |0 \rangle,
    \label{eq:coherent-state-def}
\end{align}
where $\phi$ is required to satisfy $\phi^\dag \phi = \mathbf{1}_k$. 
It can be verified that these satisfy the following properties:
\begin{align}
    \langle \phi | \phi' \rangle
    &= \det (\phi^\dag \phi')^\Nc, \\
    \langle \phi | S_\alpha^\beta | \phi \rangle 
    &= \frac{\Nc}{k} (\phi  \phi^\dag)^\beta{}^\xdag_\alpha \\
    \tint d\Omega_\phi |\phi \rangle \langle \phi | &= \mathsf{I}. 
\end{align}

Let us now compute the coherent state path integral representation for a single spin $(k,p)$ with Hamiltonian $H = J^\alpha_\beta S_\alpha^\beta $. The transition amplitude for infinitesimal imaginary time interval $\delta \tau$ in the coherent state basis is given by
\begin{align}
    & \langle \phi(\tau) | ( \mathsf{I} - \delta \tau H ) |\phi(\tau - \delta \tau) \rangle \nonumber \\
    &\quad= \langle \phi(\tau) | \phi(\tau - \delta \tau) \rangle - \delta \tau \langle \phi(\tau) | H |\phi(\tau - \delta \tau) \rangle.
\end{align}
The second term simply gives $\frac{p}{k} J^\alpha_\beta (\phi  \phi^\dag)^\beta {}^\xdag_\alpha$, while the first term gives
\begin{align}
    \langle \phi(\tau) | \phi(\tau - \delta \tau) \rangle 
    &= \det \lbrack \phi(\tau)^\dag \phi(\tau- \delta \tau)\rbrack^\Nc\nonumber \\
    &\approx \det \lbrack\mathbf{1} - \delta \tau \,\phi(\tau)^\dagger \del_\tau \phi(\tau) \rbrack^\Nc \nonumber \\
    &\approx \lbrack 1 - \delta \tau \Tr \phi(\tau)^\dagger \del_\tau \phi(\tau) \rbrack^p \nonumber \\
    &\approx 1 - p \delta \tau \Tr \phi(\tau)^\dagger \del_\tau \phi(\tau). 
\end{align}
We thus find the coherent state path integral $\Tr e^{-\beta H} = \int e^{-S} \mathcal{D} \phi \mathcal{D} \phi^\dag$ where
\begin{align}
    L = p \Tr (\phi^\dag \del_\tau \phi) 
    + \frac{p}{k} J^\alpha_\beta  (\phi  \phi^\dag)^\beta {}^\xdag_\alpha . 
    \label{eq:single-spin-action}
\end{align}

We will also need the coherent states for the conjugate representation $(\overline{k,p})$. Thus, we introduce an additional set of $N\times k$ harmonic oscillators,
\begin{align}
    [\Bar{a}^{\dag i \alpha}, \Bar{a}_{\beta j}] = \delta^\alpha_\beta \delta^i_j,
\end{align}
in terms of which $\SU(N)$ spin operators may be given by
\begin{align}
    \Sbar^\alpha_\beta = - \Bar{a}^{\dag i \alpha} \Bar{a}_{\beta i}. 
\end{align}
The corresponding coherent states $|\Bar{\chi}\rangle$ are defined analogously to \eqref{eq:coherent-state-def}
and they satisfy the properties
\begin{align}
    \langle \Bar{\chi} | \Bar{\chi}' \rangle
    &= \det (\Bar{\chi}^\dag \Bar{\chi}')^\Nc, \\
    \langle \Bar{\chi} | \Sbar^\alpha_\beta | \Bar{\chi} \rangle 
    &= - \frac{p}{k} (\chi \chi^\dag)^\alpha{}^\xdag_\beta, \\
    \tint d\Omega_\chi |\Bar{\chi} \rangle \langle \Bar{\chi} | &= \mathsf{I} \ \ \text{on $(\overline{k,p})$}.
\end{align}
The coherent state path integral for the single spin Hamiltonian $H = J_\alpha^\beta \Sbar^\alpha_\beta $ is obtained in the same way as before, and one finds the Euclidean Lagrangian
\begin{align}
    L = -p  \Tr (\chi^\dag \del_\tau \chi) 
    - \frac{p}{k} J_\alpha^\beta (\chi \chi^\dag)^\alpha {}^\xdag_\beta. 
    \label{eq:conjugate-single-spin-action}
\end{align}
Note that this is simply the negative of \eqref{eq:single-spin-action}. 

\subsection*{Large distance effective field theory}

Returning now to the spin chain \eqref{eq:spin-chain-h}, we obtain via the coherent state path integral the Euclidean Lagrangian
\begin{align}
    L 
    &=
    \tsum_n \Tr (
    p \phi_n{}^{\!\dag} \del_\tau \phi_n 
    - p \chi_n{}^{\!\dag} \del_\tau \chi_n)
    \nonumber \\
    &\;\;\;\; - \frac{p^2}{k^2} J_- \tsum_n \lvert \phi_n{}^{\!\dag} \chi_n \rvert^2
    - \frac{p^2}{k^2} J_+ \tsum_n \lvert\chi_n{}^{\!\dag}  \phi_{n+1}\rvert^2
\end{align}
We note that for $J_+ = J_-$ this Lagrangian is invariant under the offsite transformation
\begin{align}
    \phi_n \mapsto \chi_n{}^{\!*}, \ \chi_n \mapsto \phi_{n+1}{}^{\!*},
    \label{eq:spin-chain-Csym}
\end{align}
which we shall later interpret as charge conjugation.
To obtain the large-distance effective field theory, we expand about a slowly varying N\'eel configuration by writing
\begin{align}
    \phi_{n} &= \phi(x), \\ 
    \chi_{n} &= \bigg[ 1 + \frac{1}{k} \lvert \epsilon(x)\rvert^2 \bigg]^{\frac{1}{2}} \phi(x) + \epsilon(x).
\end{align}
Here, $\phi(x)$ is taken to be a slowly varying background, $\epsilon(x)$ a rapidly varying fluctuation, and we have introduced the continuum coordinate $x = an$, $a$ being the lattice spacing.
Inserting this into the Lagrangian, and keeping only terms at most second order in $\epsilon$ and $\partial$, we find
\begin{align}
    a \mathcal{L} 
    &= 
    \frac{p^2}{k^2} a^2 J_+\lvert D_x \phi\rvert^2 + \frac{p^2}{k^2}(J_- + J_+)  \lvert \epsilon\rvert^2 \nonumber \\
    &\;\;\;\;
    - \Tr \bigg[ \epsilon^\dag \bigg( \frac{p^2}{k^2} a J_+D_x \phi + p D_\tau \phi \bigg)\bigg]
    \nonumber \\
    &\;\;\;\;
    - \Tr \bigg[\bigg(\frac{p^2}{k^2} a J_+D_x \phi^\dag-p D_\tau \phi^\dag \bigg) \epsilon \bigg],
\end{align}
where $D_\mu \phi \equiv \partial_\mu \phi - \phi(\phi^\dag \partial_\mu \phi)$.
Integrating out the fluctuation field $\epsilon$ yields
\begin{align}
    \mathcal{L}
    &= a \frac{p^2}{k^2} \frac{J_-J_+}{J_-+ J_+} \lvert D_x \phi\rvert^2 
    + \frac{1}{a}\frac{k^2}{J_-+J_+} \lvert D_\tau \phi\rvert^2 \nonumber \\
    &\;\;\;\;+ p \frac{J_+}{J_-+ J_+} \epsilon^{\mu \nu} \Tr(D_\mu \phi^\dag D_\nu \phi).
\end{align}
This is the continuum Lagrangian of the $\Gr_k(N)$ model with coupling given by 
\begin{align}
    \frac{1}{g^2} = p\frac{(J_-J_+)^{\frac{1}{2}}}{J_-+J_+} = \tfrac{1}{2}p\lbrack(1-\gamma)(1+\gamma)\rbrack^{\frac{1}{2}}
\end{align}
with a $\theta$ parameter given by
\begin{align}
    \theta = 2 \pi p \frac{J_+}{J_- + J_+} = \pi p (1 + \gamma ),
\end{align}
where we have introduced the staggering parameter $\gamma$ by $J_+/J_- = (1+\gamma)/(1-\gamma)$.
For $\gamma=0$, we recover the result $\theta = \pi p$ obtained in Ref.~\cite{read_features_1989}. Thus the presence of the lattice symmetry \eqref{eq:spin-chain-Csym} exactly coincides with the presence of charge conjugation symmetry in the effective field theory. 


\end{document}